\documentclass[conference]{IEEEtran}
\IEEEoverridecommandlockouts
\usepackage[square, comma, sort&compress, numbers]{natbib}
\usepackage{amsmath,amssymb,amsfonts}
\usepackage{graphicx}

\usepackage[caption=false,font=footnotesize]{subfig}
\usepackage{textcomp}
\graphicspath{{figures/}} 
 
\usepackage[margin=0.75in,footskip=0.25in]{geometry}

\usepackage[ruled,vlined]{algorithm2e}
\usepackage{setspace}
\SetAlgoSkip{}
\SetAlgoInsideSkip{}
\SetAlFnt{\footnotesize}
\SetAlCapFnt{\footnotesize}
\SetAlCapNameFnt{\footnotesize}
\SetKwRepeat{Do}{do}{while}
    
\begin{document}

\title{ADMM Enabled Hybrid Precoding in Wideband Distributed Phased Arrays Based MIMO Systems}

%\author{\IEEEauthorblockN{Yu Zhang\textsuperscript{1,2},~\IEEEmembership{Student Member,~IEEE}, 
%Yiming Huo\textsuperscript{2},~\IEEEmembership{Member,~IEEE}, Jinlong Zhan\textsuperscript{2}, \\ Dongming Wang\textsuperscript{1},~\IEEEmembership{Member,~IEEE}, Xiaodai Dong\textsuperscript{2}~\IEEEmembership{Senior Member,~IEEE}, and Xiaohu You\textsuperscript{1}~\IEEEmembership{Fellow,~IEEE}}
%\IEEEauthorblockA{
%\textsuperscript{1}National Mobile Communications Research Laboratory, Southeast University, Nanjing 210096, China \\
%Email: \{yuzhang, wangdm, xhyu\}@seu.edu.cn  \\
%\textsuperscript{2}Department of Electrical and Computer Engineering, University of Victoria, Victoria BC V8W 3P6, Canada \\
%Email: \{yuzhang3, ymhuo, jinlongzhan\}@uvic.ca and xdong@ece.uvic.ca}
%}
\author{\IEEEauthorblockN{Yu Zhang\textsuperscript{1,2},
Yiming Huo\textsuperscript{2}, Jinlong Zhan\textsuperscript{2},  Dongming Wang\textsuperscript{1}, Xiaodai Dong\textsuperscript{2}, and Xiaohu You\textsuperscript{1}}
\IEEEauthorblockA{
\textsuperscript{1}National Mobile Communications Research Laboratory, Southeast University, Nanjing 210096, China \\
Email: \{yuzhang, wangdm, xhyu\}@seu.edu.cn  \\
\textsuperscript{2}Department of Electrical and Computer Engineering, University of Victoria, Victoria BC V8W 3P6, Canada \\
Email: \{yuzhang3, ymhuo, jinlongzhan\}@uvic.ca and xdong@ece.uvic.ca}
}

\maketitle

\begin{abstract}
Distributed phased arrays based multiple-input multiple-output (DPA-MIMO) is a recently proposed highly reconfigurable architecture enabling both spatial multiplexing and beamforming in millimeter-wave (mmWave) systems. In this work, we focus on coping with the hybrid precoding for the wideband DPA-MIMO system with orthogonal frequency division multiplexing (OFDM) modulation. More specifically, we propose an alternating direction method of multipliers (ADMM) enabled hybrid precoding approach based on an alternating optimization framework, abbreviated to ADMM-AltMin, for such cooperative array-of-subarrays structures. Simulation results show that the proposed ADMM-AltMin method achieves favourable performance with practical quantization of phase shifters taken into account.
\end{abstract}


% -----------------------------------------------
%                     SECTION  
% ----------------------------------------------- 
\section{Introduction}
Millimeter-wave (mmWave) communications can significantly boost the channel capacity and enhance the quality of service (QoS)  of next-generation wireless networks due to large radio frequency (RF) bandwidth in the less crowded spectrum typically from 30 GHz to 300 GHz, e.g., \cite{pi2011introduction,yue2017demonstration,li2018mmwave}. Despite more severe pathloss and atmospheric absorption in mmWave bands compared to the conventional sub-6 GHz frequencies, more antennas can be accommodated into a relatively small hardware area to form large-scale antenna arrays at transceiver ends \cite{huo2014wideband}. These massive antenna arrays can facilitate large beamforming gains to combat the undesired propagation characteristics of mmWave channels, e.g., \cite{akdeniz2014millimeter,huang2018novel,rusek2013scaling}. Compared to the fully-digital beamforming schemes with one dedicated RF chain for each antenna element, hybrid beamforming schemes, implemented by connecting only several RF chains to all the antennas through phase shifters, can achieve almost the same spectral efficiency performance but with lower power consumption, e.g., \cite{liang2014low,el2014spatially}.

\begin{figure}[!t]
\centering
\includegraphics[width=2.7in]{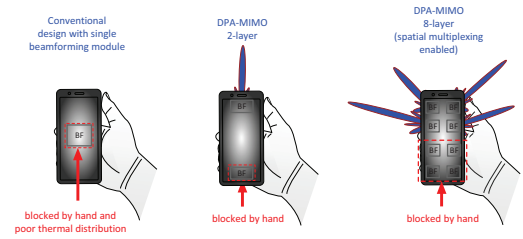}
\caption{DPA-MIMO architecture facilitating next-generation user equipment design, with a comparison with conventional mmWave beamforming design \cite{huo2018cellular}.}
\label{Fig_DPA_UE}
\vspace{-10pt}
\end{figure}
In order to realize a design tradeoff of both diversity and spatial multiplexing gains in such hybrid analog-digital structures, a natural strategy is to deploy multiple seperated subarray modules at a transmitter (TX) and/or a receiver (RX), where each module can perform independent analog beam steering  \cite{singh2015feasibility}. By considering the practical design principles and constraints of antennas, circuits and systems, and user equipment (UE) designs, a novel reconfigurable architecture, called distributed phased arrays based MIMO (DPA-MIMO), has recently been developed to provide flexible beamforming strategies and high data throughput for fifth generation (5G) cellular communications \cite{huo20175g,zhang2019channel}. Such array of distributed subarrays structure and its practical implementation architecture present more appealing application solutions including but not limited to, e.g., WiFi and cellular co-enabling cost-effective architecture that facilitates 5G super carrier aggregation (`Super-CA') among sub-6 GHz bands and mmWave bands \cite{huo2018cellular}; multi-beam multi-stream mmWave communications for distributed and multi-layer unmanned aerial vehicle (UAV) networks \cite{huo2018millimeter,huo2019distributed}; satellite communications, etc. For example, Fig.{~\ref{Fig_DPA_UE}} further demonstrates a vivid illustration of how this DPA architecture can solve and mitigate human hand/body blockage issue for the 5G and beyond user equipment compared to the conventional mmWave design with single beamforming module.

Most of the published literature about subarray-based hybrid beamforming designs, e.g., {\cite{singh2015feasibility,gao2016energy,lota20175g}} mainly focus on the narrowband channel model. Practical mmWave MIMO systems, however, need to deal with frequency-selective channels due to the large signal bandwidth, e.g.,  \cite{alkhateeb2016frequency,sohrabi2017hybrid}. The challenge of wideband hybrid precoding designs lies in that a common RF precoder should be shared by all subcarriers while baseband precoding is adopted for individual subcarrier. A Gram-Schmidt based frequency-selective hybrid precoding is proposed based on the designed codebooks of limited feedback for the fully-connected structure \cite{alkhateeb2016frequency}.  Furthermore, an alternating optimization based hybrid precoder design is applied to the wideband partially-connected structure \cite{yu2016alternating}. However,                                                                                                                                                                                 with a fixed RF precoder, a non-convex baseband precoder design problem is transformed into a semidefinite relaxation (SDR) problem that consumes high complexity. Recently, alternating direction method of multipliers (ADMM) has received intense attention due to its good behavior in dealing with some non-convex problems in machine learning areas \cite{boyd2011distributed}. Moreover, ADMM has been applied in the mmWave systems for multicast transmission \cite{huang2017low} and energy-efficient hybrid analog-digital transceiver designs \cite{tsinos2017energy}.

In this work, we design hybrid precoding for the wideband DPA-MIMO system implemented using orthogonal frequency division multiplexing (OFDM) modulation, where spatially seperated and distributed, but still cooperative subarrays experience independent mmWave channels. Based on the alternating optimization framework \cite{ni2017near}, we develop an ADMM enabled wideband hybrid precoding method for such array-of-subarrays structures. In details, we first resort to ADMM to solve the non-convex baseband precoder design problem with the fixed RF precoder, and then derive the closed-form expression for the optimal RF precoder with the obtained baseband precoder. Finally, numerical results demonstrate that, with less computation required, the proposed hybrid precoding method can achieve satisfactory performance compared to the SDR based approach \cite{yu2016alternating}. 

\textbf{Notation}: Bold uppercase $\mathbf{A}$ (bold lowercase $\mathbf{a}$) denotes a matrix (a vector). $\left[ \mathbf{A}\right]_{i,j}$  and $\left[ \mathbf{A}\right]_{i,:}$ denotes the $\left( i,j \right)$th element and the $i$th row, respectively. ${\mathrm{Tr}}\lbrace \cdot  \rbrace$ stands for the matrix trace operation.  ${{\mathbf{I}}_N}$ denotes a $N \times N$ dimensional identity matrix. ${\left(  \cdot  \right)^H}$, ${\left(  \cdot  \right)^T}$ , ${\left(  \cdot  \right)^ * }$ and ${\left(  \cdot  \right)^{-1} }$ stand for the conjugate transpose, transpose, conjugate and inverse, respectively. ${\mathcal{CN}}\left( \boldsymbol{\mu } , \mathbf{R} \right)$ denotes the complex Gaussian distribution with mean $\boldsymbol{\mu }$ and covariance matrix $\mathbf{R}$. $\Re \left \{ \cdot  \right \}$ and $\Im  \left \{ \cdot  \right \}$ denote the real part and imaginary part of a matrix, respectively. $\left \lceil \cdot  \right \rceil$ is the operation of rounding up to an integer and  ${\mathbb{E}}\lbrace\cdot  \rbrace$ is the expectation operator. Finally, $\mathrm{blkdiag}\left \{ \mathbf{A}_1,\cdots ,\mathbf{A}_N \right \}$ denotes  a block diagonal matrix whose diagonal entries are given by $\mathbf{A}_1,\cdots ,\mathbf{A}_N$.
  
% -----------------------------------------------
%                     SECTION  
% ----------------------------------------------- 
\section{System Model}
\begin{figure}[!t]
\centering
\includegraphics[width=2.7in]{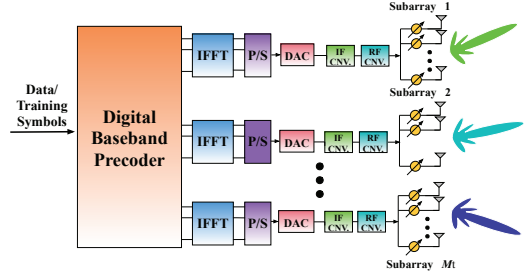}
\caption{Wideband DPA-MIMO system at the transmitter end.}
\label{Fig_DPA_BS}
\vspace{-5pt}
\end{figure}

Consider a wideband DPA-MIMO system where a TX with $M_t$ subarrays, each of which has one RF chain, transmits $N_s$ data streams to a fully-digital RX with $N_r$ antennas, as shown in Fig.{~\ref{Fig_DPA_BS}}. The total number of antennas at the TX  is denoted by $N_t^{\rm tot}$.  Note that $N_t^{\rm tot} = M_t \times N_t^{\rm sub}$, where $N_t^{\rm sub}$ is the number of antennas at each subarray. In our setup, each subarray is a uniform linear array (ULA) and all the subarrays are arranged in the same axis (straight line) at the TX.  The edge-to-edge distance of the neighboring linear subarrays is much larger than the antenna element distance within a subarray.

\subsection{Channel Model}
Considering the limited scattering characteristics of the mmWave propagation environment, e.g.,  \cite{alkhateeb2016frequency,park2017dynamic,yu2016alternating}, we adopt a geometric channel with $N_{cl}$ scattering clusters and $N_{ray}$ rays within each cluster. Furthermore, we assume that the channels of different subarrays at the TX are independent due to large spacing (e.g., more than 2 free-space wavelengths) between neighbouring subarrays \cite{lin2015adaptive}. The channel matrix of the $k$th subcarrier between the $m$th TX subarray and the RX is represented as
\begin{equation}
\label{equ_channel_subarray}
\mathbf{H}_{m}\left [ k \right ]=\gamma  \sum_{i=1}^{N_{cl}}\sum_{\ell=1}^{N_{ray}}\alpha _{i \ell}^{\left [ m \right ]}\mathbf{a}_r\left ( \varphi _{i,\ell}^{\left [ m \right ]} \right )\mathbf{a}_t^H\left (\theta^{\left [ m \right ]} _{i\ell} \right )e^{-j\frac{2\pi i k}{K}},  
\end{equation}
where $K$ is the number of subcarriers and $\gamma = \sqrt{\frac{N_t^{\mathrm{sub}}N_r^{\mathrm{sub}}}{N_{cl}N_{ray}}}$ denotes the normalized factor. $\alpha _{i \ell}^{\left [ m \right ]} \sim \mathcal{CN}\left ( 0,1 \right )$ is the normalized complex gain of the $\ell$th ray in the $i$th cluster. $\varphi_{i\ell}^{\left [ m \right ]}$ and $\theta _{i\ell}^{\left [ m \right ]}$ represent angles of departure (AoDs) and angles of arrival (AoAs) of the $m$th subarray, respectively. The antenna array response vector in a ULA configuration with $N$ antennas is
\begin{equation}
\label{equ_array_response}
\mathbf{a}\left ( \phi  \right ) = \frac{1}{\sqrt{N}}\left [ 1, e^{j\frac{2\pi }{\lambda }d_e \sin\phi } ,\cdots ,e^{j\frac{2\pi }{\lambda }\left ( N-1 \right )d_e \sin\phi } \right ]^T ,
\end{equation}
where $\lambda$ is the carrier wavelength and $d_e$ denotes antenna element spacing within each subarray. Thus, we can define the entire channel matrix of the $k$th subcarrier between all TX subarrays and the RX using
\begin{equation}
\label{equ_entire_channel}
\mathbf{H}\left [ k \right ] = \left [ \mathbf{H}_1 \left [ k \right ],\cdots ,\mathbf{H}_{M_t}\left [ k \right ] \right ].
\end{equation}

\subsection{Transmission}
As shown in Fig.{~\ref{Fig_DPA_BS}}, the TX precodes the $N_s$ data symbols $\mathbf{s}\left[ k\right]$ at each subcarrier $k = 1,\cdots ,K$ with  a digital precoding matrix  $\mathbf{F}_{\mathrm{BB}} \left [ k \right ]\in \mathbb{C}^{M_t\times N_s}$, and then transforms the symbol vector prepared for each TX subarray to the time domain with parallel $K$-point inverse fast Fourier transform (IFFT) operations. After adding cyclic prefixes (CPs), an RF precoding matrix $\mathbf{F}_{\mathrm{RF}} \in \mathbb{C}^{N_t^{\mathrm{tot}}\times M_t}$ is employed to obtain the final transmitted signal at the $k$th subcarrier given by
\begin{equation}
\label{equ_x_k}
\mathbf{x}\left [ k \right ] =\mathbf{F}_{\mathrm{RF}}\mathbf{F}_{\mathrm{BB}}\left [ k \right ] \mathbf{s}\left [ k \right ],
\end{equation}
where $\mathbf{s}\left [ k \right ] \in \mathbb{C}^{N_s \times 1}$ is the transmitted vector of the $k$th subcarrier which satisfies $\mathbb{E}\left \{ \mathbf{s}\left [ k \right ]\mathbf{s}\left [ k \right ]^H \right \}=\frac{P}{N_s}\mathbf{I}_{N_s}$ with a total transmit power budget per subcarrier $P$.

Assuming perfect symbol time and frequency offset synchronization at the RX, the CP is first removed to form the received sequence. Then, the symbols are transformed to the frequency domain through an FFT operation. Furthermore, with a fully-digital combining matrix $\mathbf{W}\left [ k \right ] \in \mathbb{C}^{N_r\times N_s}$ for each subcarrier, the final received signal of the $k$th subcarrier is given by 
\begin{equation}
\label{equ_y}
\mathbf{y} \left [ k \right ] = \mathbf{W}\left [ k \right ] ^H\mathbf{H}\left [ k \right ] \mathbf{x}\left [ k \right ] +  \mathbf{W}\left [ k \right ] ^H \mathbf{z}\left [ k \right ],  
\end{equation}
where $\mathbf{z}\left [ k \right ] \sim \mathcal{CN}\left ( 0, \sigma _{\mathrm{z}}^2 \mathbf{I}_{N_r^\mathrm{tot}} \right )$ is an additive white Gaussian noise vector for the $k$th subcarrier with a variance $\sigma _{\mathrm{z}}^2$. Note that the RF precoding matrix has a block diagonal structure, i.e., $\mathbf{F}_\mathrm{RF}= \mathrm{blkdiag}\left \{ \mathbf{f}_{\mathrm{RF}}^{\left [ 1 \right ]},\cdots ,\mathbf{f}_{\mathrm{RF}}^{\left [ M_t \right ]} \right \}$ where $\mathbf{f}_{\mathrm{RF}}^{\left [ m \right ]} \in \mathbb{C}^{N_t^\mathrm{sub}\times 1}$ represents the RF precoder for the $m$th TX subarray.

\subsection{Problem Formulation}
Following the matrix approximation approach, e.g., \cite{el2014spatially,lee2014af,yu2016alternating}, we aim to  minimize the sum of the Frobenius norms of the differences between the optimal digital precoder and the multiplication of the RF precoder and the per-subcarrier baseband precoder. The corresponding mathematical formulation is written as 
\begin{subequations}
\label{alternate_optimization_TX_design}
\begin{alignat}{3}
& \underset{ 
\begin{scriptsize}
\mathbf{F}_{\mathrm{RF}},  \left \{ \mathbf{F}_{\mathrm{BB}}\left [ k \right ]\right \}_{k=1}^K
 \end{scriptsize}
}{\text{min}}
\label{alternate_optimization_TX_design_objective}
&\quad & \sum_{k=1}^{K}\left \| \mathbf{F}_{\mathrm{opt}}\left [ k \right ] - \mathbf{F}_{\mathrm{RF}}\mathbf{F}_{\mathrm{BB}}\left [ k \right ] \right \|_F^2 \\
& \qquad \;\;\,\text{s.t.}
\label{alternate_optimization_TX_design_constraint_power}
& & \left \| \mathbf{F}_{\mathrm{RF}} \mathbf{F}_{\mathrm{BB}}\left[k\right] \right \|_{F}^{2} = N_s \\
\label{alternate_optimization_TX_design_constraint_RF}
&&& \left | \left [ \mathbf{F}_{\mathrm{RF}}  \right ]_{i,j} \right |  = \frac{1}{\sqrt{N_{t}^{\mathrm{sub}}}}, \, \forall \left ( i,j \right ) \in \mathcal{F}_{t},
\end{alignat}
\end{subequations}
where  $\mathbf{F}_{\mathrm{BB}}\left [ k \right ]$ stands for the baseband precoder of the $k$th subcarrier and $\mathcal{F}_{t}$ denotes the index set of nonzero elements of the RF precoder. The optimal precoder of the $k$th subcarrier is given by $\mathbf{F}_{\mathrm{opt}}\left [ k \right ]  = \mathbf{V}\left [ k \right ] \boldsymbol{\Sigma }\left [ k \right ]^{1/2}$ in which $\mathbf{V}\left [ k \right ] $ is composed of the $N_s$ singular vectors corresponding to the $N_s$ largest singular values of $\mathbf{H}\left [ k \right ]$ and $\boldsymbol{\Sigma }\left [ k \right ]$ is the water-filling power allocation matrix of the $k$th subcarrier. This problem is non-convex due to the constant amplitude constraint of the phase shifters \eqref{alternate_optimization_TX_design_constraint_RF} and the coupling of the RF and baseband precoders \eqref{alternate_optimization_TX_design_objective} \cite{ni2017near}. Note that we only focus on the hybrid precoding design in the following parts, while the hybird combing design can be performed by the same procedure if given the DPA-MIMO architecture at the RX, e.g., \cite{lee2014af,alkhateeb2016frequency}.
% -----------------------------------------------
%                     SECTION  
% ----------------------------------------------- \\
\section{ADMM Enabled Hybrid Precoding}
%\vspace{-5pt}
  In this section, we give a brief introduction to ADMM and then present an ADMM enabled hybrid precoding algorithm based on the well-known alternating optimization framework.
%\vspace{-15pt}
\subsection{ADMM for Non-convex Problems}
ADMM is an algorithm that is intended to combine the decomposability of dual ascent with the superior convergence of the method of multipliers \cite{boyd2011distributed}. Consider a constrained optimization problem
\begin{subequations}
\begin{alignat}{2}
& \underset{ 
\begin{scriptsize}
 \mathbf{x} \in \mathbb{R}^{N\times 1}
 \end{scriptsize}
}{\text{min}}
&\quad &  f(\mathbf{x}) \\
& \;\;\;\;  \text{s.t.}
&&  \mathbf{x} \in \mathcal{C},
\end{alignat}
\end{subequations}
where $f(\mathbf{x})$ is convex, but $\mathcal{C}$ is non-convex. Then, a new variable $\mathbf{y}$ is introduced to split the objective function in the above problem which is converted to
\begin{subequations}
\begin{alignat}{2}
& \underset{ 
\begin{scriptsize}
 \mathbf{x} \in \mathbb{R}^{N\times 1}
 \end{scriptsize}
}{\text{min}}
&\quad &  f(\mathbf{x}) + I_{\mathcal{C}}\left ( \mathbf y \right ) \\
& \;\;\;\;  \text{s.t.}
&&  \mathbf{x} -\mathbf{y} = \mathbf{0},
\end{alignat}
\end{subequations}
where the indicator function $I _{\mathcal{C}}\left( \mathbf{x} \right )$ associated with set $\mathcal{C}$ is defined as
\begin{equation}
\label{equ_indicator_fun}
I_{\mathcal{C}}\left ( \mathbf{x} \right ) =\left\{
\begin{matrix}
0,\;\;\,\mathbf{x} \in \mathcal{C}\\ 
\infty ,\; \mathbf{x}\notin  \mathcal{C}
\end{matrix}.
\right.
\end{equation}

Using the scaled ADMM algorithm \cite{boyd2011distributed},  the problem can be iteratively solved by the following equations 
\begin{subequations}
\label{admm_iterate}
\begin{align}
{\mathbf{x}}_{i+1} &=   \underset{ {\mathbf{x} \in \mathbb{R}^{N\times 1}
}}{\arg  \min} \; f\left ( {\mathbf{x}} \right )+\left ( \rho / 2 \right ) \left \| {\mathbf{x}}-\mathbf{y}_i+\boldsymbol{\nu}_i \right \|_2^2, \\
\mathbf{y}_{i+1} &= \Pi _{\mathcal{C}}\left ({\mathbf{x}}_{i+1} + \boldsymbol{\nu}_i  \right ), \\
\label{equ_iterate_nu}
\boldsymbol{\nu}_{i+1} &= \boldsymbol{\nu}_i  + {\mathbf{x}}_{i+1} - \mathbf{y}_{i+1},
\end{align}
\end{subequations}
where $\rho $ is a parameter that places a penalty on violations of primal feasibility, $\boldsymbol{\nu}_i $ is the dual variable, and $\Pi _{\mathcal{C}}\left ( \cdot  \right )$ is projection onto set ${\mathcal{C}}$. 
%The projection can be accomplished by solving the following convex problem 
%\begin{subequations}
%\label{problem_projection}
%\begin{alignat}{2}
%& {\text{min}}
%&\quad & \left \| \mathbf{z}- \mathbf{x} \right \|_2 \\
%& \; \text{s.t.}
%&& \mathbf{z} \in {\mathcal{C}}.
%\end{alignat}
%\end{subequations}
%\vspace{-17pt}
\subsection{Hybrid Precoder Design}
%\vspace{-5pt}
\subsubsection{Baseband Precoder Design}
Based on the alternating optimization framework, we first design the baseband precoder with the fixed RF precoder. Therefore, the problem of \eqref{alternate_optimization_TX_design} is decoupled into per-subcarrier baseband precoder design, which is written as 
\begin{subequations}
\label{problem_baseband_precoder}
\begin{alignat}{2}
& \underset{ 
\begin{scriptsize}
\mathbf{F}_{\mathrm{BB}}\left [ k \right ]
 \end{scriptsize}
}{\text{min}}
\label{problem_baseband_precoder_objective}
&\quad & \left \| \mathbf{F}_{\mathrm{opt}}\left [ k \right ] - \mathbf{F}_{\mathrm{RF}}\mathbf{F}_{\mathrm{BB}}\left [ k \right ] \right \|_F^2 \\
\label{problem_baseband_precoder_constraint}
& \;\; \text{s.t.}
& & \left \| \mathbf{F}_{\mathrm{RF}} \mathbf{F}_{\mathrm{BB}}\left[k\right] \right \|_{F}^{2} = P.
\end{alignat}
\end{subequations}
The power constraint \eqref{problem_baseband_precoder_constraint} can be further simplified to 
\begin{equation}
\label{equ_power_constraint}
\left \| \mathbf{F}_{\mathrm{RF}} \mathbf{F}_{\mathrm{BB}}\left[k\right] \right \|_{F}^{2} = \left \| \mathbf{F}_{\mathrm{BB}}\left[k\right] \right \|_{F}^{2} = P.
\end{equation}
Applying a vectorization operation to the objective function \eqref{problem_baseband_precoder_objective}, we can obtain
\begin{equation}
\label{equ_vec_Fopt}
\begin{aligned}
\eqref{problem_baseband_precoder_objective} &=  \left \| \mathrm{vec} \left \{ \mathbf{F}_{\mathrm{RF}} \mathbf{F}_{\mathrm{BB}}\left [ k \right ]  \right \}  - \mathrm{vec}\left \{ \mathbf{F}_{\mathrm{opt}}\left [ k \right ]  \right \}\right \|_F^2 \\
&= \left \| \left ( \mathbf{I}_{N_s}\otimes \mathbf{F}_{\mathrm{RF}}  \right )\mathrm{vec} \left \{  \mathbf{F}_{\mathrm{BB}}\left [ k \right ] \right \}  - \mathrm{vec} \left \{  \mathbf{F}_{\mathrm{opt}}\left [ k \right ] \right \} \right \|_2^2
\end{aligned}
\end{equation}
For notational simplicity, we denote $c = P$, $\mathbf{A} = \mathbf{I}_{N_s}\otimes \mathbf{F}_{\mathrm{RF}}$, $\mathbf{x} = \mathrm{vec} \left \{  \mathbf{F}_{\mathrm{BB}}\left [ k \right ] \right \}  $ and $\mathbf{b} = \mathrm{vec} \left \{  \mathbf{F}_{\mathrm{opt}}\left [ k \right ] \right \}$.  Thus, the problem of \eqref{problem_baseband_precoder} is converted into the following form 
\begin{subequations}
\label{problem_hQCQP_complex}
\begin{alignat}{2}
& \underset{ 
\begin{scriptsize}
\mathbf{x} \in \mathbb{C}^{M_tN_s\times 1}
 \end{scriptsize}
}{\text{min}}
& \quad & \left \| \mathbf{A}\mathbf{x} -\mathbf{b} \right \|_2^2  \\
& \;\;\;\;\;\; \text{s.t.}
& &  \left \| \mathbf{x}  \right \|_2^2 =c.
\end{alignat}
\end{subequations}
Although the objective function of \eqref{problem_hQCQP_complex} is convex, the feasible region however is not. Furthermore, in order to avoid undefined derivative of the real function on complex variables, we formulate the problem of \eqref{problem_hQCQP_complex} into real domain as
\begin{subequations}
\label{problem_hQCQP_real}
\begin{alignat}{2}
& \underset{ 
\begin{scriptsize}
\bar{\mathbf{x}} \in \mathbb{R}^{2M_tN_s\times 1}
 \end{scriptsize}
}{\text{min}}
&\quad & g \left( \bar{\mathbf{x}} \right)\\
& \;\;\;\;\;\;\, \text{s.t.}
& &  \left \| \bar{\mathbf{x}}  \right \|_2^2 =c,
\end{alignat}
\end{subequations}
in which $\bar{\mathbf{x}} = \left [ \Re\left \{ \mathbf{x} \right \}^T, \Im\left \{ \mathbf{x} \right \}^T\right ]^T$, and 
%\vspace{-5pt}
\begin{equation}
\label{equ_g}
g \left( \bar{\mathbf{x}} \right)=  \left \| \mathbf{A}_1\bar{\mathbf{x}}-\Re \left \{  \mathbf{b}\right \}   \right \|_2^2 + \left \| \mathbf{A}_2\bar{\mathbf{x}}-\Im  \left \{  \mathbf{b}\right \}   \right \|_2^2,
\end{equation}
where  $\mathbf{A}_1 = \left [ \Re \left \{ \mathbf{A} \right \}, -\Im \left \{ \mathbf{A} \right \}\right ]$ and $\mathbf{A}_2 = \left [ \Im \left \{ \mathbf{A} \right \}, \Re \left \{ \mathbf{A} \right \} \right ]$.

Via doing such transformation, we are able to apply scaled ADMM to the problem of \eqref{problem_hQCQP_real} by the following iterations
\begin{equation}
\label{admm_iterate_x_bar}
\begin{aligned}
\bar{\mathbf{x}}_{i+1} &= \underset {\bar{\mathbf{x}} \in \mathbb{R}^{2M_tN_s\times 1}}{\arg \min} g \left( \bar{\mathbf{x}} \right)+\left ( \rho / 2 \right ) \left \| \bar{\mathbf{x}}-\mathbf{y}_i+\boldsymbol{\nu}_i \right \|_2^2 \\
& = \frac{2 \mathbf{A}_1^T \Re \left \{ \mathbf{b}  \right \}  + 2\mathbf{A}_2^T \Im \left \{ \mathbf{b}  \right \}  +\rho \left ( \mathbf{y}_i - \bar{\mathbf{x}}_i  \right )}{2+\rho },
\end{aligned}
\end{equation}
\begin{equation}
\label{admm_iterate_y}
\mathbf{y}_{i+1}  = \Pi _{\mathcal{C}}\left (\bar{\mathbf{x}}_{i+1} + \boldsymbol{\nu}_i  \right ) 
\overset{\left ( a \right )}{=} \sqrt{c}\frac{\bar{\mathbf{x}}_{i+1} + \boldsymbol{\nu}_i }{\left \| \bar{\mathbf{x}}_{i+1} + \boldsymbol{\nu}_i  \right \|_2},
\end{equation}
where ${\mathcal{C}} = \left \{ \bar{\mathbf{x}} \, | \left \| \bar{\mathbf{x}}  \right \|_2^2 =c \right \}$ and $\left ( a \right )$ follows from a geometric interpretation, i.e., projection of a point $\left ( \bar{\mathbf{x}}_{i+1} + \boldsymbol{\nu}_i  \right )$ onto a multi-dimentional sphere $\mathcal{C}$ \cite{lu2018distributed}. Then, these iterations keep running until the following termination criteria are met
\begin{subequations}
\begin{align}
\left \| \bar{\mathbf{x}}_{i+1} - {\mathbf{y}}_{i+1}  \right \|_2 & < \epsilon _{\mathrm{p}}, \\
\left \| \rho \left (\mathbf{y}_{i+1} - \mathbf{y}_{i}  \right )\right \|_2 & < \epsilon _{\mathrm{d}}.
\end{align}
\end{subequations}
where $\epsilon _{\mathrm{p}}$ and $\epsilon _{\mathrm{d}}$ denote predefined tolerances for the primal residual and the dual residual, respectively.

\subsubsection{RF Precoder Design}
With the obtained baseband precoder, we can submit it into the original problem of \eqref{alternate_optimization_TX_design}, which leads to the following form
\begin{equation}
\label{problem_opt_phase}
\underset{\vartheta _{il}}{\min} \sum_{k=1}^{K} \left \| \left [ \mathbf{F}_{\mathrm{opt}}\left [ k \right ] \right ]_{i,:} - \frac{1}{\sqrt{N_t^{\mathrm{sub}}}}e^{j\vartheta _{il}}\left [ \mathbf{F}_{\mathrm{BB}}\left [ k \right ] \right ]_{l,:} \right \|_2^2,
\end{equation}
where $\vartheta _{il}$ is the phase of the $\left ( i,l \right )$th element of $\mathbf{F}_{\mathrm{RF}}$, $1\leq i \leq N_t^{\mathrm{tot}}$ and $l=\left \lceil \frac{i}{N_t^{\mathrm{sub}}} \right \rceil$. In fact, this problem has a closed-form solution \cite{yu2016alternating} as
\begin{equation}
\label{equ_opt_phase}
\vartheta _{il}^\star =  \measuredangle \left ( \sum_{k=1}^{K} \left [ \mathbf{F}_{\mathrm{opt}}\left [ k \right ] \right ]_{i,:} \left [ \mathbf{F}_{\mathrm{BB}}\left [ k \right ] \right ]_{l,:}^H  \right ),
\end{equation}
where $\measuredangle \left( \cdot \right)$ extracts the corresponding phases of the elements.

In the previous analysis, the phase of each entry of $\mathbf{F}_{\mathrm{RF}}$  is assumed to be a continuous value. However, this assumption contradicts the practical designs and applications with finite resolution phase shifters where the phase shifting is quantized to limited bits \cite{liang2014low}. Therefore, we can easily obtain every element of the optimal quantized RF precoder through
\begin{equation}
\label{equ_opt_RF}
\left (\left [ \mathbf{F}_{\mathrm{RF}} \right ]_{i,l}  \right )^\star  =\frac{1}{\sqrt{N_t^{\mathrm{sub}}}}e^{j\mathcal{Q} \left \{ \vartheta _{il}^\star\right \}},
\end{equation}
where $\mathcal{Q} \left \{ \cdot  \right \} $ is a function that quantizes the continuous phase to its nearest neighbouring distrete phase values based on closest Euclidean distance \cite{li2018mmwave}. The proposed hybrid precoding algorithm is summarized in Algorithm{~\ref{alg_analog_precoder_SIC}} which facilitates a distributed realization.  
\begin{algorithm}[!h]
\caption{ADMM Based Hybrid Precoder Design}
\label{alg_analog_precoder_SIC}
\SetKwInOut{Input}{Input}
\SetKwInOut{Output}{Output}
\SetKwFor{While}{while}{}{}%
\Input{$ \left \{ \mathbf{F}_{\mathrm{opt}} \left [ k \right ]  \right \}_{k=1}^K $.} 
Initialize $\left [ \mathbf{F}_{\mathrm{RF}} \right ]_{i,j} = 1/\sqrt{N_t^{\mathrm{sub}}} ,\, \forall \left ( i,j \right ) \in \mathcal{F}_{t}$ and $n = 0$.\\
\Repeat{some termination criterion is satisfied}{
1. Fix $\mathbf{F}_{\mathrm{RF}}^{\left ( n \right )}$ and solve $\mathbf{F}_{\mathrm{BB}}^{\left ( n\right )}$ using \eqref{admm_iterate_x_bar}, \eqref{admm_iterate_y} and \eqref{equ_iterate_nu}.\\
2. Fix $\mathbf{F}_{\mathrm{BB}}^{\left ( n \right )}$ and update $\mathbf{F}_{\mathrm{RF}}^{\left ( n+1\right )}$ by \eqref{equ_opt_RF}.\\
3. $n = n+1$.}
\Output{$\left \{ \mathbf{F}_{\mathrm{BB}}^\star \left [ k \right ]  \right \}_{k=1}^K$, $\mathbf{F}_{\mathrm{RF}}^\star $.}
\end{algorithm}
% -----------------------------------------------
%                     SECTION  
% ----------------------------------------------- 
\section{Simulation Results}
%\begin{figure}[!t]
%\centering
%\includegraphics[width=3.5in]{Fig_SE_SNR_Cmp.eps}
%\caption{Average spectral efficiency vs SNR of different hybrid precoding methods with infinite resolution of phase shifters given $N_t^{\mathrm{sub}} = 8$, $M_t = 4$, $N_r = 8$, $K = 32$ and $N_s = 2$ .}
%\label{Fig_SE_SNR_Cmp}
%\end{figure}
%
%\begin{figure}[!t]
%\centering
%\includegraphics[width=3.5in]{Fig_SE_SNR_BitNum.eps}
%\caption{Average spectral efficiency vs SNR under different resolution of  phase shifters given $N_t^{\mathrm{sub}} = 16$, $M_t = 6$, $N_r = 16$,  $K = 64$ and $N_s = 3$ .}
%\label{Fig_SE_SNR_BitNum}
%\end{figure}
%
%\begin{figure}[!t]
%\centering
%\includegraphics[width=3.5in]{Fig_SE_SNR_CSI.eps}
%\caption{Impact of imperfect CSI on ADMM based hybrid precoding with different number of antennas at both TX and RX ends given $M_t = 8$, $K = 64$ and $N_s =4$.}
%\label{Fig_SE_CSI}
%\end{figure}
\begin{figure*}[!t]
\centering
\subfloat[]{\includegraphics[width=2.2in]{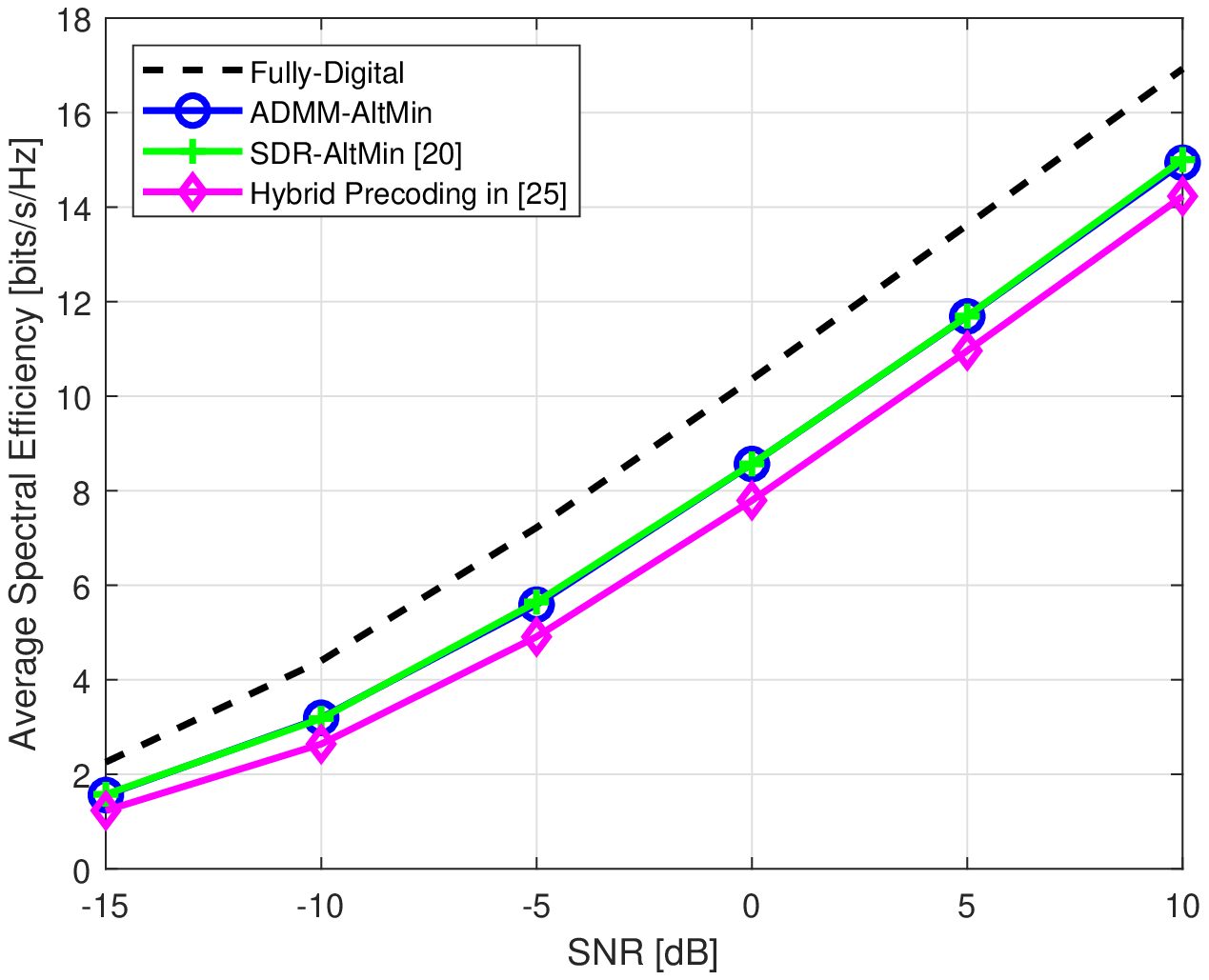}
\label{Fig_SE_SNR_Cmp}}
%\hspace{-0.5in}
\subfloat[]{\includegraphics[width=2.2in]{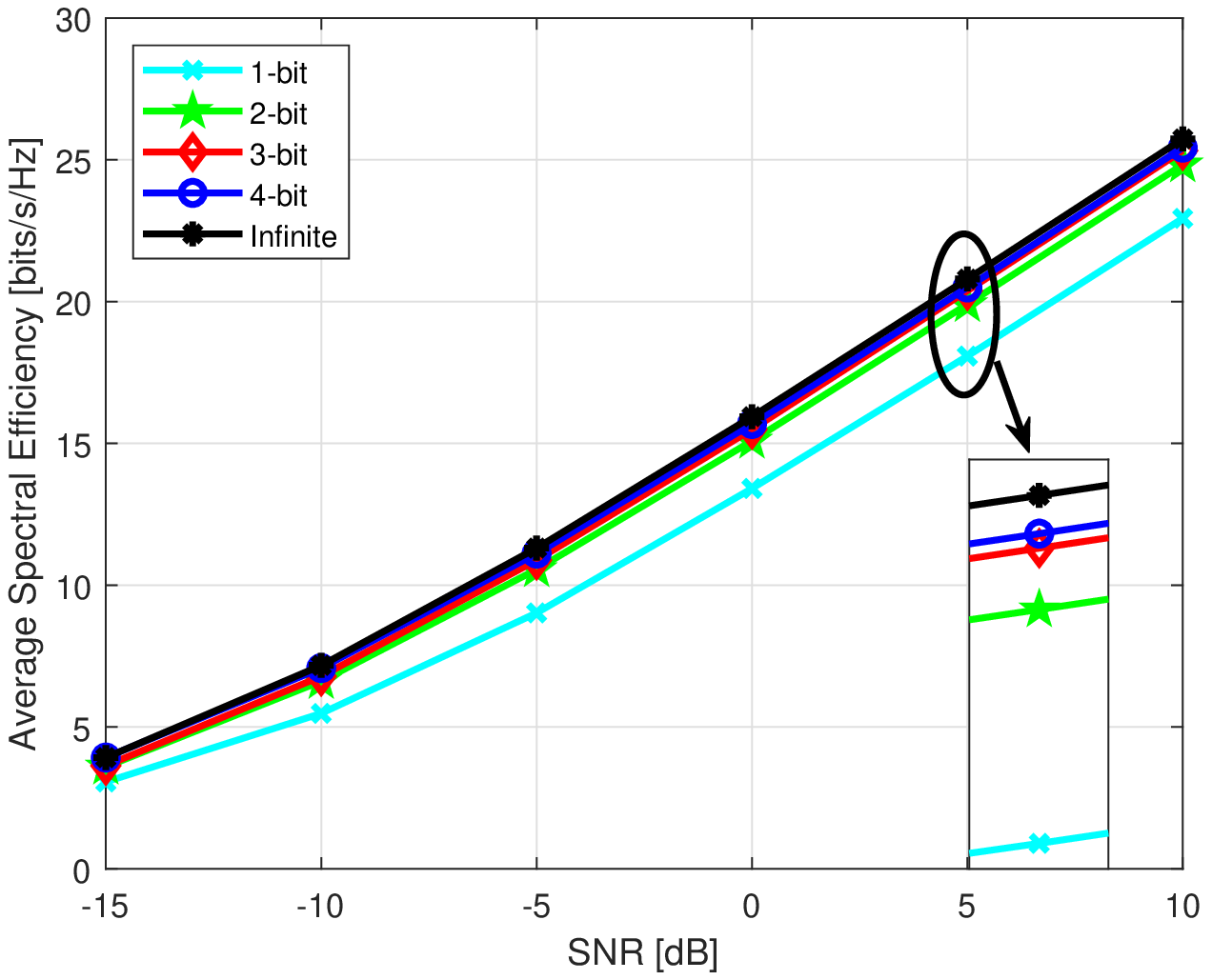}
\label{Fig_SE_SNR_BitNum}}
%\hspace{-0.5in}
\subfloat[]{\includegraphics[width=2.2in]{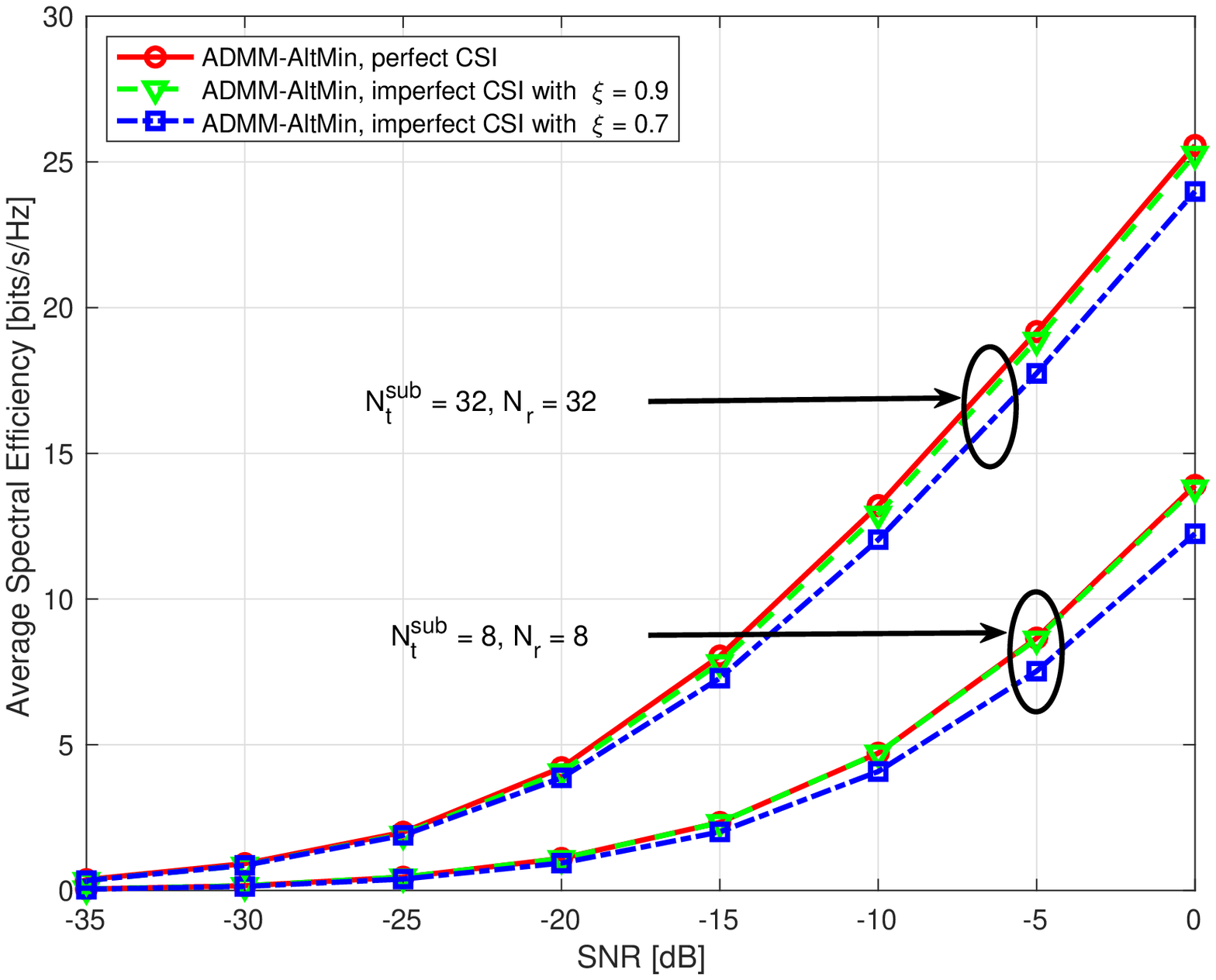}
\label{Fig_SE_SNR_CSI}}
\caption{(a) Average spectral efficiency vs SNR of different hybrid precoding methods with infinite resolution of phase shifters given $N_t^{\mathrm{sub}} = 8$, $M_t = 4$, $N_r = 8$, $K = 32$ and $N_s = 2$. (b) Average spectral efficiency vs SNR under different resolution of  phase shifters given $N_t^{\mathrm{sub}} = 16$, $M_t = 6$, $N_r = 16$,  $K = 64$ and $N_s = 3$. (c) Impact of imperfect CSI on ADMM based hybrid precoding with different number of antennas at both TX and RX ends given $M_t = 8$, $K = 64$ and $N_s =4$.}
\vspace{-5pt}
\end{figure*} 
In this section, we evaluate the performance of the proposed hybrid precoding algorithm. In the simulations, we consider the ULA configuration with half-wavelength (free space) antenna element spacing within each subarray at the TX{\footnote{Note that there is large distance separation among subarrays and the channel parameters for each subarray are generated independently in this paper.}}. Furthermore, the parameters of mmWave channels are set as, $N_{cl} = 5$ clusters and $N_{ray}$ = 10 rays \cite{akdeniz2014millimeter}. The AoDs and AoAs are generated by the Laplacian distribution with uniformly distributed mean angles over $\left [ 0,2\pi  \right )$ and each cluster's angular spread of 10 degrees \cite{alkhateeb2016frequency}. The total transmit power budget per subcarrier is set as $P = N_s$ and the average spectral efficiency is plotted versus the signal-to-noise-ratio (SNR) per-subcarrier that is defined as $\mathrm{SNR} = \frac{P}{\sigma_{\mathrm{z}} ^2} $. Within the iterations of ADMM, we set $\rho = 1$ and $\epsilon _{\mathrm{p}} = \epsilon _{\mathrm{d}} = 10^{-6}$.

As illustrated in Fig.{~\ref{Fig_SE_SNR_Cmp}}, we compare the performance of the proposed ADMM-AltMin scheme with the SDR-AltMin algorithm  in \cite{yu2016alternating} and the hybrid precoding method based on the conventional architecture of fixed subarrays in \cite{park2017dynamic} with infinite resolution phase shifters when $N_t^{\mathrm{sub}} = 8$, $M_t = 4$, $N_r = 8$, $K = 32$ and $N_s = 2$. It is shown that the proposed scheme achieves the same performance as the SDR-AltMin algorithm. For the baseband precoder design, the SDR-AltMin algorithm needs to solve an SDR problem with the per-iteration complexity of $\mathcal{O}\left( \max {{\left\{ {{M}_{t}}{{N}_{s}}+1,3 \right\}}^{4}}{{\left( {{M}_{t}}{{N}_{s}}+1 \right)}^{{1}/{2}\;}}\log \left( {1}/{\varepsilon }\; \right) \right)$ where $\varepsilon$ is a predefined accuracy parameter \cite{luo2010semidefinite}, while the proposed scheme performs simple real-value computation with that of $\mathcal{O}\left( {M_t^2N_s^4} \right)$. Moreover, the ADMM-AltMin method significantly outperforms the  algorithm in \cite{park2017dynamic}.

Fig.{~\ref{Fig_SE_SNR_BitNum}} shows the impact of finite resolution phase shifters on the proposed scheme when $N_t^{\mathrm{sub}} = 16$, $M_t = 6$, $N_r = 16$,  $K = 64$ and $N_s = 3$. As the number of quantization levels of phase shifters changes from 1-bit to 4-bit, the performance of average spectral efficiency approximates to the phase shifters with infinite resolution. This fact proves that the proposed method is suitable and ideal for practical communication applications where hardware resources are limited, particularly in a UE device \cite{huo20175g}.

On the other hand, Fig.{~\ref{Fig_SE_SNR_CSI}} evaluates the impact of imperfect channel state information (CSI) on the proposed algorithm when $M_t = 8$, $K = 64$ and $N_s =4$. The estimated channel matrix of the $k$th subcarrier can be modeled as \cite{rusek2013scaling}
\begin{equation}
\label{equ_H_xi}
\hat{\mathbf{H}}\left [ k \right ] = \xi \mathbf{H}\left [ k \right ] + \sqrt{1-\xi^2}  \mathbf{E}\left [ k \right ]
\end{equation}
where $\xi \in \left [ 0,1 \right ]$ presents the CSI accuracy, and $\mathbf{E} \left [ k \right ]$ is the error matrix for the $k$th subcarrier with entries following the i.i.d. $\mathcal{CN}\left ( 0,1 \right )$. We can observe that the average spectral efficiency becomes higher with larger number of antenna elements at both TX and RX ends, and the proposed scheme is able to handle the scenarios of imperfect CSI.

% -----------------------------------------------
%                     SECTION  
% ----------------------------------------------- 
\section{Conclusion}
In this paper, we present the design of ADMM based hybrid precoders for the wideband DPA-MIMO system on top of the alternating optimization framework. Through conducting numerical analysis and comparison, the proposed ADMM-AltMin method can achieve the same spectral efficiency as the higher-complexity SDR-AltMin method in \cite{yu2016alternating}. Furthermore, the proposed scheme only needs 4-bit phase shifters for cost-efficient system implementation in practical 5G and Beyond applications, while still providing good immunity to the imperfect CSI. For future work, it would be interesting to focus on broadband multiuser (MU) downlink transmission with both the BS and the UE on top of the DPA-MIMO architecture.

% -----------------------------------------------
%                     SECTION  
% ----------------------------------------------- 
\section{Acknowledgements}
{
%\small
This work was supported in part by the National Key Special Program under Grant 2018ZX03001008-002, the National Natural Science Foundation of China (NSFC) under Grant 61871122, Grant 61801168, and Grant 61571120, and the Six Talent Peaks Project in Jiangsu Province.
}

% -----------------------------------------------
%                     SECTION  
% ----------------------------------------------- 
% references
{
\scriptsize
\bibliographystyle{IEEEtran}
\bibliography{IEEEabrv,mybibfile}
}

\end{document}